\documentclass[12pt]{article}
\usepackage{epsfig}
\usepackage{amsmath}
\usepackage{latexsym}
\usepackage{amssymb}
\usepackage{hyperref}

\def\Bra#1{\mathinner{\langle{#1}|}}
\def\Ket#1{\mathinner{|{#1}\rangle}}
\def\braket#1{\mathinner{\langle{#1}\rangle}}

\def\ket#1{\left|#1\right>}
{\catcode`\|=\active 
  \gdef\Braket#1{\left<\mathcode`\|"8000\let|\bravert {#1}\right>}}
\def\bravert{\egroup\,\vrule\,\bgroup}
\newcommand{\be}{\begin{eqnarray}}
\newcommand{\ee}{\end{eqnarray}}
\newcommand{\bea}{\begin{eqnarray}}
\newcommand{\eea}{\end{eqnarray}}
\newcommand{\ben}{\begin{equation}}
\newcommand{\een}{\end{equation}}

\newcommand{\Rhat}{\widehat R}

\newcommand{\nn}{\nonumber}

\numberwithin{equation}{section}

\begin{document}

\begin{titlepage}

\begin{flushright}
CALT-68-2500\\
hep-th/0405172
\end{flushright}
\vspace{15 mm}
\begin{center}
{\huge On the Integrability of String Theory \\
in $AdS_5\times S^5$  }
\end{center}
\vspace{12 mm}
\begin{center}
{\large 
Ian Swanson}\\
\vspace{3mm}
California Institute of Technology\\
Pasadena, CA 91125, USA 
\end{center}
\vspace{5 mm}
\begin{center}
{\large Abstract}
\end{center}
\noindent

Integrability occupies an increasingly important role in direct tests of the
AdS/CFT correspondence.  Integrable structures have appeared 
in both planar ${\cal N}=4$ super Yang-Mills theory and type IIB superstring 
theory on $AdS_5\times S^5$.  A generalized statement of the AdS/CFT 
conjecture has therefore emerged in which, in addition to string energies
corresponding to gauge theory anomalous dimensions, an infinite
tower of higher charges on each side of the duality should also 
be equated.  Demonstrations of this larger equivalence have been successful in certain 
regimes.  To test this correspondence in a more stringent
setting, the bosonic sector of the fully quantized string theory on $AdS_5\times S^5$
is expanded about the pp-wave limit to sextic order in fields, or to $O(1/J^2)$,
where $J$ is the (large) angular momentum of string states boosted along an equatorial
geodesic in the $S^5$ subspace.  To avoid issues of renormalization, the analysis
is restricted to zeroth order in the modified 't~Hooft coupling
where consistency conditions demand that integrability be realized.
The string theory, however, fails to meet these conditions.   
This signals a potential problem with higher-order corrections in the large-$J$ 
expansion around the pp-wave limit.

\vspace{1cm}
\begin{flushleft}
\today
\end{flushleft}
\end{titlepage}
\newpage
\section{Introduction}

Studies of the AdS/CFT correspondence have made impressive strides in
recent years.  The complications involved in quantizing type
IIB superstring theory on $AdS_5\times S^5$ in the presence of background
Ramond-Ramond fields were circumvented to some extent by Metsaev, who
was able to show that in a particular kinematic limit the string
theory in this background becomes free \cite{Metsaev:2001bj,Metsaev:2002re}.  
In this limit, states are boosted
along a null geodesic in the $S^5$ subspace, and the geometry
is reduced to a pp-wave \cite{Blau:2001ne,Blau:2002dy,Blau:2002mw}.  On the gauge theory side
a corresponding limit of planar ${\cal N}=4$ supersymmetric Yang-Mills (SYM)
theory with $SU(N_c)$ gauge group was identified by Berenstein, Maldacena and 
Nastase, where the anomalous dimensions of a class of single-trace SYM operators 
with large $U(1)$ component of the $SU(4)$ $R$-charge were matched to the string 
theory energy spectrum on the pp-wave \cite{Berenstein:2002jq}.  

The dimensions of ${\cal N}=4$ SYM operators in the planar (large $N_c$) limit are
perturbative in the 't Hooft coupling $\lambda = g_{YM}^2 N_c$ and
are analytic functions of the scalar $R$-charge, which is dual to 
the $S^5$ angular momentum $J$ of states in the string theory.
Since string energies are exact in the string dual of the 't 
Hooft coupling (known as the modified 't Hooft coupling 
$\lambda' = g_{YM}^2 N_c / J^2$), the resulting duality landscape is 
one in which agreement is obtained in the overlap between the large-$J$, 
small-$\lambda'$ limit of the string theory, and the large-$R$, 
small-$\lambda$ limit of the gauge theory.  The correspondence can be 
probed at a much deeper level by including higher $\lambda$
loop corrections to the gauge theory and higher-order $1/J$ corrections
to the string theory.  This program has been pursued is a series of
recent studies 
(see, eg.~\cite{Parnachev:2002kk,Callan:2003xr,Callan:2004uv,3IMP,
Beisert:2003tq,Beisert:2003jj,Beisert:2003jb,Beisert:2003ea,
Beisert:2003ys,Arutyunov:2003rg,Serban:2004jf,Arutyunov:2004xy,Beisert:2004hm}).

Finite-$J$ corrections to the string spectrum can be interpreted
as interaction perturbations to the worldsheet theory arising from
finite-radius curvature corrections to the pp-wave geometry.  
This curvature expansion has thus far been carried out 
in the fully supersymmetric theory to $O(1/J)$
(or, in terms of the spacetime radius $\Rhat$, to $O(1/\Rhat^2)$)
for two and three-impurity string states \cite{Callan:2003xr,Callan:2004uv,3IMP}.  
On the gauge theory side, the problem of computing anomalous dimensions of operators
has undergone a striking simplification:  in various subsectors of the theory
the dilatation operator can be mapped to the Hamiltonian of an integrable 
spin-chain.
The results have confirmed expectations 
of the AdS/CFT correspondence to $O(1/J)$ in 
the curvature expansion and $O(\lambda^2)$ in the gauge loop expansion.
At the present time, however, there is a perplexing disagreement
between each side of the correspondence at three loops in $\lambda$
\cite{Callan:2003xr,Callan:2004uv,3IMP,Beisert:2003ea,Serban:2004jf}.

This disagreement aside, 
the emergence of integrable structures on each side of the duality
has come to play a central role in explorations of the AdS/CFT 
mechanism.  (For a review of current progress, 
see, eg.~\cite{Arutyunov:2004xy,Beisert:2004hm}.)  
This direction of study was launched
by Minahan and Zarembo in \cite{Minahan:2002ve}, wherein it was shown that at one loop
in $\lambda$ the anomalous dimension matrix of SYM operators in an
$SU(2)$ subsector of the theory is precisely the Hamiltonian of the
so-called ${\rm XXX}_{1/2}$ integrable Heisenberg spin chain.
Integrability in these systems can be proved by
showing that the $R$ matrix associated with the dilatation operator
satisfies the Yang-Baxter equation.  Anomalous dimensions are then typically
computed by means of the Bethe ansatz technique \cite{Faddeev:1996iy,Faddeev:ph}.
(In \cite{Beisert:2003yb}, this application was generalized 
to the full ${\it PSU}(2,2|4)$ super spin chain at one loop in $\lambda$.)  
The integrable structure generates an infinite tower of higher 
commuting (local) charges, denoted by $\{Q_{k}\}$ (with $k=0,\dots,\infty$), 
where $Q_0$ is a cyclic shift in the trace, $Q_1$ is the anomalous dilatation 
operator and higher $Q_k$ represent a series of hidden Abelian charges (most of which have not
been linked to the known symmetries of ${\cal N}=4$ SYM theory \cite{Arutyunov:2004xy}).  
In certain closed subsectors of the theory this structure has been shown to
remain intact to the three-loop level \cite{Beisert:2003tq,Beisert:2003ys}, 
and an obvious implication is that the theory may in fact be integrable at all loops, 
generating a tower of hidden charges that are exact in $\lambda$. 
One may therefore promote the charges $\{Q_{k}\}$ to
analytic functions of $\lambda$ where, in the notation of \cite{Beisert:2004hm},
\be 
Q_k(\lambda) = \sum_{j=0}^\infty \lambda^{j} Q_{k,2j}\ .
\ee

It should also be noted that a similar direction of investigation
has been pursued for a class of non-local (non-Abelian) charges generated by Yangian
structures on each side of the correspondence \cite{Mandal:2002fs,Bena:2003wd,Alday:2003zb}.  
Successful contact was made at one loop between corresponding infinite-dimensional
non-Abelian symmetry algebras on either side of the duality in
\cite{Dolan:2003uh,Dolan:2004ps}.  Here, however, we will primarily 
be concerned with the Abelian sector of the integrable structure.

The conjectured all-loop integrability is supported by 
the fact that certain integrable 
structures emerge from the classical string sigma model.
Classical solutions corresponding to rigid strings moving in the 
$S^5$ subspace \cite{Gubser:2002tv,Frolov:2003qc,Frolov:2003xy,Arutyunov:2003uj} 
are characterized by Neumann integrable systems, 
and naturally give rise to an infinite tower of hidden commuting charges
that are non-perturbative in $\lambda$ 
\cite{Arutyunov:2003rg,Serban:2004jf,Engquist:2004bx}.
At one loop, string energies in these semiclassical systems were matched to the
$SU(2)$ Bethe ansatz results for the spin chain in \cite{Beisert:2003ea,Beisert:2003xu}.
Beyond one loop, where the spin chain is characterized by non-nearest-neighbor interactions, 
the dilatation operator acquires long-range terms that are not immediately soluble in
terms of the Bethe ansatz approach.  This challenge was surmounted in \cite{Serban:2004jf},
where the Inozemtsev long-range spin chain was employed to develop a long-range
Bethe ansatz, and string predictions were matched to two loops.
Moreover, the higher commuting charges in the gauge theory have been shown to match 
the corresponding classical string charges to two loops in $\lambda$ by comparing the long-range
Bethe ansatz with the classical Bethe equation associated with the string sigma model 
\cite{Arutyunov:2004xy,Kazakov:2004qf}.
This matching can be seen as a consequence of the fact that, to two-loop order, 
the classical string action is identical to the effective
two-dimensional action of the coherent-state vector field describing the 
corresponding spin chain system \cite{Kruczenski:2003gt}.
(This analysis was extended to higher order in the semiclassical 
$1/J$ expansion in \cite{Kruczenski:2004kw}.)
The conjectured exact equivalence of the full tower of commuting charges 
in the string and gauge theories 
(and the corresponding duality relationship in their respective couplings) 
can be taken as a generalized version of the AdS/CFT correspondence.
In direct tests, however, this dramatic agreement begins to break down at three-loop 
order \cite{Serban:2004jf,Arutyunov:2004xy}.  

The purpose of the present study is to move beyond the semiclassical
limit of the string theory and test integrability in the fully
quantized theory at higher orders in the string background curvature
expansion.  We will focus on a particular conserved charge $Q_2(\lambda)$ and
its string counterpart $Q_2^{\rm string}(\lambda')$, restricted to two 
closed bosonic subsectors in each theory.  In the CFT, these protected subsectors appear as 
$SL(2)$ and $SU(2)$ bosonic sectors that cannot mix with any other states in the 
theory, to all orders in $\lambda$.  In the string theory 
these subsectors correspond to bosonic symmetric-traceless states that decouple 
in either the $SO(4)$ subspace descending from $AdS_5$ ($SO(4)_{AdS}$) 
or in the $SO(4)$ descending from the $S^5$ ($SO(4)_{S^5}$).

In the gauge theory, the conserved charge 
$Q_2(\lambda)$ was studied to two loops in the closed $SU(2)$ bosonic 
subsector in \cite{Beisert:2003tq}, where it was shown that, in addition to 
commuting with the dilatation operator $Q_1(\lambda)$, 
$Q_2(\lambda)$ anticommutes with a parity operator $P$ which acts 
on single-trace operators by inverting the order of fields 
within the trace.  In addition, $Q_2(\lambda)$ can also be shown to connect
operators of opposite parity, so that 
the existence of $Q_2(\lambda)$ gives rise to degenerate
pairs of operators (under $Q_1(\lambda)$) connected by $P$.
Conversely, parity degeneracy in the spectrum of $Q_1(\lambda)$ at a given order in the loop 
expansion implies that $Q_2(\lambda)$ can be computed explicitly to that order.  
This degeneracy was originally used to fix the form of 
the three-loop dilatation operator $Q_{1,4}$ 
in the $SU(2)$ closed subsector which, without
assuming integrability (and hence parity degeneracy), was only fixed up 
to two free coefficients \cite{Beisert:2003tq,Beisert:2003jb}.\footnote{To be precise,
one of these coefficients is fixed by demanding proper scaling in the BMN
limit, and the other is fixed by parity degeneracy.}
The status of $Q_{1,4}$  has since improved:  it can be fixed by independent 
symmetry arguments \cite{Beisert:2003ys} and, in accordance with the expectations of 
integrability, the theory in this subsector does indeed exhibit parity degeneracy 
to three loops.  Beyond this, the requirements of BMN scaling and parity degeneracy
fix the form of the dilatation operator to {\it five} loops.  $Q_{1,6}$ and $Q_{1,8}$,
however, have yet to be independently confirmed by symmetry arguments alone.

As demonstrated in \cite{3IMP}, the string theory version of parity 
corresponds to exchanging left and right-moving modes on the worldsheet.
Since the agreement between string energies at $O(1/J)$ and anomalous dimensions fails
at three loops in $\lambda$, it may have been reasonable to expect the breakdown of parity
degeneracy in the string theory at this order.  According to the results in \cite{3IMP}, 
however, this is in fact not the case:  $Q_2^{\rm string}(\lambda')$ commutes with the string 
Hamiltonian and parity degeneracy persists to $O(1/J)$ and to all loops in 
the gauge coupling.\footnote{It is necessary to consider at least three worldsheet 
impurities to study these aspects of parity in the string theory.  The same statement 
in the gauge theory says that one requires at least three impurities in the trace
to admit states with distinct parity in the same group representation \cite{Beisert:2003tq}.}
The disagreement between gauge and string theory at three loops in $\lambda$ 
is due to an overall shift in the string energy spectrum that preserves 
this particular facet of the integrable structure.  

When treated as a constraint, integrability in the gauge theory is an extremely restrictive
requirement \cite{Beisert:2004hm}. 
Given the established disagreement with string theory at 
three loops, it may not be surprising to see a breakdown of 
integrability on the string side of the duality at some order 
in the $1/J$ expansion.\footnote{Note that the three-loop disagreement is 
by no means inextricably tied to the survival of integrability.}
At $O(1/J^2)$, the most basic question is of course whether there is any
agreement between string energies and gauge theory anomalous dimensions.
At this order, however, the string theory becomes 
subject to several renormalization issues, and these interesting yet complicated 
problems will be reserved for a subsequent paper.  
The subject of the present study will be a more immediate test on $Q_2^{\rm string}(\lambda')$.  
In particular, we will expand the bosonic sector of the string theory 
to sextic order in fields and test whether $Q_2^{\rm string}(\lambda')$ is still conserved 
at $O(1/J^2)$.  

In section 2, curvature corrections to the Green-Schwarz superstring
action in the pp-wave limit of $AdS_5\times S^5$ will be reviewed.
Employing the techniques described in \cite{Callan:2004uv}, corrections to the 
action will be extended to $O(1/J^2)$ (or $O(1/\Rhat^4)$ in terms of
the curvature radius) in the bosonic sector of the theory. In section
3, matrix elements of the resulting curvature corrections will be computed
for two classes of closed, three-impurity string states.
It will be shown that the basic expectations from gauge theory integrability
are not met at this order in the expansion.  We conclude with a 
discussion of this problem and of future directions of study.

\section{Higher curvature corrections to the Hamiltonian}

As described in \cite{Callan:2003xr,Callan:2004uv}, 
the pp-wave limit of $AdS_5\times S^5$ can acquire
finite-radius curvature corrections which lead to interaction
perturbations to the free worldsheet theory.
Here, corrections to the pp-wave Hamiltonian are arranged according to the expansion
\be
H = \sum_{k=0} \frac{H^{(k)}}{\Rhat^{2k}}\ ,
\ee
where $H^{(0)}$ is the zeroth-order theory in the full Penrose limit
of the geometry, and $\Rhat$ is the spacetime radius.  
We will employ the same form of the $AdS_5\times S^5$ metric used in
\cite{Callan:2003xr,Callan:2004uv}:
\begin{equation}
\label{metric}
ds^2  =  \Rhat^2
\biggl[ -\left(\frac{1+ \frac{1}{4}z^2}{1-\frac{1}{4}z^2}\right)^2dt^2
        +\left(\frac{1-\frac{1}{4}y^2}{1+\frac{1}{4}y^2}\right)^2d\phi^2
    + \frac{d z_k dz_k}{(1-\frac{1}{4}z^2)^{2}}
    + \frac{dy_{k'} dy_{k'}}{(1+\frac{1}{4}y^2)^{2}} \biggr]~.
\end{equation}
The time $t$ and $\phi$ directions will be combined to form lightcone coordinates $x^\pm$,
while $x^A$ ($A=1,\dots,8$) will label eight transverse directions  
which are broken into the two $SO(4)$ subspaces noted above.
These are denoted by $z^2 = z_k z^k$, which span the $SO(4)_{AdS}$ 
(with $k=1,\dots,4$), and $y^2 = y_{k'} y^{k'}$, which span the $SO(4)_{S^5}$ subgroup
($k'=5,\dots,8$). The radius $\Rhat$ is related to the gauge theory 't 
Hooft coupling by $\Rhat^4 = \lambda (\alpha')^2$.  

For reasons described
in \cite{Callan:2004uv}, we use a particular choice of lightcone coordinates
given by
\be
\label{newcoords}
t  =  x^+\, , \qquad
\phi  =  x^+ +{x^-}\ .
\ee
In the Penrose limit, where states are boosted along an equator in the $S^5$ subspace,
$\phi$ and the transverse coordinates $z_k$ and $y_{k'}$ are rescaled according to
\be
\phi \to x^+ + \frac{x^-}{\Rhat^2} \qquad 
z_k \to \frac{z_k}{\Rhat} \qquad y_{k'} \to \frac{y_{k'}}{\Rhat}\ .
\ee
This leads to the following curvature expansion of (\ref{metric}) 
in powers of $1/\Rhat^2$ about the Penrose limit:
\be
\label{Rhatexp}
ds^2 & = & 2dx^+ dx^- - (x^A)^2 (dx^+)^2 + (dx^A)^2 
\nonumber \\
& & 	+ \frac{1}{\Rhat^2}\left[ 
	-2y^2 dx^+ dx^- + \frac{1}{2}(y^4-z^4) (dx^+)^2 + (dx^-)^2
	+\frac{1}{2}z^2 dz^2 - \frac{1}{2}y^2 dy^2 \right]
\nonumber \\
& & 	+ \frac{1}{\Rhat^4}\Bigl[
	16 y^4 dx^+ dx^- - 3 (x^A)^6 (dx^+)^2 - 16 y^2 (dx^-)^2
	+3 y^4 dy^2 + 3 z^4 dz^2 \Bigr]
\nn\\
&&	+ {O}\left({1}/{\Rhat^6}\right)\ .
\ee
The leading-order term is the lightcone metric of the pp-wave, and the
$O(1/\Rhat^2)$ curvature correction leads to the first $1/J$ correction
to the string spectrum, which is the subject of 
\cite{Callan:2003xr,Callan:2004uv,3IMP}.

The supersymmetric Green-Schwarz action describing
type IIB string theory on this background is constructed 
from the Cartan one-forms and superconnections on
the coset space 
$G/H = [ SO(4,2)\times SO(6) ]/ [SO(4,1)\times SO(5) ]$.
Here we intend to focus on the bosonic sector of the theory: 
the salient points in this study 
can be made without confronting the complications
present in the fermionic sector (specifically, these arise from the existence of second-class
constraints on the fermionic degrees of freedom).
Without fermions, the full $AdS_5 \times S^5$ Lagrangian takes the form
\begin{eqnarray}
\label{lagrangiank}
{\cal L} =  -\frac{1}{2} h^{ab} L_a^\mu L_b^\mu\ ,  
\label{LAG}
\end{eqnarray}  
where the Cartan one-forms $L_a^\mu$ are given simply by
\be
L_a^\mu = e^\mu_{\phantom{\mu}\nu}\partial_a x^\nu\ .
\ee
The indices $a,b = 0,1$ are used to denote the worldsheet
coordinates $\tau$ ($a,b=0$) and $\sigma$ ($a,b=1$).

The general lightcone gauge-fixing procedure is to eliminate
unphysical degrees of freedom by imposing the
gauge condition $x^+ = p_-\tau$ and enforcing both the 
$x^-$ equations of motion and the conformal gauge constraints in
the action.  In the present setting, these operations can be
achieved order-by-order in the large-$\Rhat$ expansion.  
The only complication is that, with the
lightcone coordinates chosen in (\ref{newcoords}), the worldsheet
metric must acquire curvature corrections to remain consistent
with the equations of motion.  To organize the calculation,
the worldsheet metric is taken to be flat at leading order 
and higher order curvature corrections are arranged according 
to the following expansion:
\be
h^{00} &=& -1 + \frac{h_{(2)}^{00}}{\Rhat^2} + \frac{h_{(4)}^{00}}{\Rhat^4}
	+ O(\Rhat^{-6})\ ,
\nn\\
h^{11} &=& 1 + \frac{h_{(2)}^{11}}{\Rhat^2} + \frac{h_{(4)}^{11}}{\Rhat^4}
	+ O(\Rhat^{-6})\ ,
\nn\\
h^{01} &=& \frac{h_{(2)}^{01}}{\Rhat^2} + \frac{h_{(4)}^{01}}{\Rhat^4}
	+ O(\Rhat^{-6})\ .
\label{hEXP}
\ee
As described in \cite{Callan:2003xr,Callan:2004uv}, this simply rewrites 
$h^{ab}$ and does not (at this stage) amount to a particular gauge choice.
Using a similar notation, 
the worldsheet derivatives of $x^-$ are expanded as
\be
\dot x^- = \dot x^-_{(0)} + \frac{\dot x^-_{(2)}}{\Rhat^2} 
	+ \frac{\dot x^-_{(4)}}{\Rhat^4} + O(\Rhat^{-6})\ ,   
\qquad
	{x'}^- = {x'}^-_{(0)} + \frac{{x'}^-_{(2)}}{\Rhat^2} 
	+ \frac{{x'}^-_{(4)}}{\Rhat^4} + O(\Rhat^{-6})\ .
\label{X-EXP}
\ee

To proceed, we construct the
basic combinations of Cartan one-forms appearing in the Lagrangian (\ref{LAG})
to $O(\Rhat^{-4})$:
\be
L_0^\mu L_0^\mu & = & 2 p_- \dot x^- - p_-^2 (x^A)^2 + (\dot x^A)^2
\nn\\
&&	+ \frac{1}{\Rhat^2}\biggl\{
	(\dot x^-)^2-2p_- y^2 \dot x^- + \frac{1}{2}(\dot z^2 z^2 - \dot y^2 y^2)
	+ \frac{p_-^2}{2}(y^4-z^4) \biggr\}
\nn\\
&&\kern-30pt	+\frac{1}{\Rhat^4}\biggl\{
	\frac{3}{16}\left[ \dot y^2 y^4 + \dot z^2 z^4 - p_-^2 (z^6 + y^6)\right]
	+  p_- y^4 \dot x^- - y^2 (\dot x^-)^2\biggr\} 
	+ O(\Rhat^{-6})\ ,
\ee
\be
\label{L1L1}
L_1^\mu L_1^\mu & = & 
	({x'}^A)^2 + \frac{1}{\Rhat^2}\biggl\{
	\frac{1}{2}({z'}^2 z^2 - {y'}^2 y^2) + ({x'}^-)^2 \biggr\}
\nn\\
&&	+\frac{1}{\Rhat^4}\biggl\{ 
	\frac{3}{16}\left[ {y'}^2 y^4 + {z'}^2 z^4 \right] - y^2 ({x'}^-)^2 \biggr\}
	+{O}(\Rhat^{-6})\ ,
\ee
\be 
\label{L0L1}
L_0^\mu L_1^\mu & = & 
	p_-{x'}^- + \dot x^A{x'}^A 
  	+ \frac{1}{\Rhat^2}\biggl\{ {x'}^- \dot x^- - p_-y^2{x'}^- 
	+ \frac{1}{2}(z^2 \dot z_k z'_k
	-y^2 \dot y_{k'} y'_{k'}) \biggr\}
\nn\\
&&	+\frac{1}{\Rhat^4}\biggl\{ \frac{3}{16}\left[ y^4 \dot y_{k'} y'_{k'}
	+z^4 \dot z_k z'_k \right] + \frac{1}{2} p_- y^4 {x'}^- - y^2{x'}^-\dot x^-
	\biggr\}
	+{\cal O}(\Rhat^{-6})~.
\ee
The Virasoro constraints are obtained by varying the Lagrangian 
with respect to the worldsheet metric $h^{ab}$:
\be
T_{ab} = L_a^\mu L_b^\mu - \frac{1}{2}h_{ab}h^{cd}L_c^\mu L_d^\mu = 0\ .
\label{VC}
\ee
These constraints will be used in conjunction with the $x^-$ equations of motion
to solve for higher-order corrections to $h^{ab}$.
The $x^-$ equations of motion can be satisfied by setting the following variations 
to zero:
\be
\frac{\delta{\cal L}}{\delta\dot x^-} & = &
	p_- + \frac{1}{\Rhat^2}
	\left[ \dot x^- - p_- (y^2 - h_{(2)}^{00}) \right]
\nn\\
&&	+\frac{1}{\Rhat^4}\left[
	p_-\left( h_{(2)}^{00} y^2 - h_{(4)}^{00}
	+\frac{1}{2}y^4\right) - \dot x^- (h_{(2)}^{00} - y^2)
	- h_{(2)}^{01} {x'}^- \right] + O(\Rhat^{-6})\ ,
\nn\\
\frac{\delta{\cal L}}{\delta {x'}^-} & = &
	- \frac{1}{\Rhat^2}\left[{x'}^- + p_- h_{(2)}^{01} \right]
\nn\\
&&	+\frac{1}{\Rhat^4}\left[
	p_-\left( h_{(2)}^{01} y^2 - h_{(4)}^{01}\right)
	- h_{(2)}^{01}\dot x^-
	+{x'}^- \left( y^2 - h_{(2)}^{11} \right) \right]
	+ O(\Rhat^{-6})\ .
\label{X-EOM}
\ee
In terms of the transverse 
$SO(4)_{AdS}\times SO(4)_{S^5}$ coordinates, the expansion of $x^-$ 
(\ref{X-EXP}) to $O(\Rhat^{-2})$ is
\be
\dot x^-_{(0)}  =  \frac{p_-}{2}(x^A)^2
	-\frac{1}{2p_-}\left[(\dot x^A)^2+({x'}^A)^2\right]\ ,
\qquad
{x'}^-_{(0)}  =  -\frac{1}{p_-}\dot x^A {x'}^A\ ,
\ee
\be
\dot x^-_{(2)} & = & \frac{1}{8 p_-^3}\biggl\{
	-4(\dot x^A {a'}^A)^2 
	- \left[ (\dot x^A)^2 - 3({x'}^A)^2 \right]\left[
	(\dot x^A)^2 +({x'}^A)^2 \right]
\nn\\
&&	+ 2\left[
	y^2 \left( 2 {y'}^2 + {z'}^2 - \dot z^2 \right)
	+z^2 \left(\dot y^2 - {y'}^2 - 2{z'}^2 \right)
	\right]
	+p_-^4 \left[(x^A)^2\right]^2 \biggr\}\ ,
\nn\\
{x'}^-_{(2)} & = & \frac{1}{2p_-^3}\biggl\{
	p_-^2\left( z^2 \dot y_{k'} y'_{k'} -y^2 \dot z_k z'_k \right)
	-\left( \dot x^A {x'}^A\right)\left[
	(\dot x^A)^2 - ({x'}^A)^2 \right] \biggr\}\ .
\ee
At $O(\Rhat^{-4})$ we only need to solve for $\dot x^-$ (ie.~$x^-_{(4)}$) to eliminate all
instances of $x^-$ from the Lagrangian.  We find
\be
\dot x^-_{(4)} & = & -\frac{1}{32 p_-^5}\biggl\{
	2\Bigl[
	\dot y^6 - \dot y^4({y'}^2 - 3\dot z^2 )
	+4 (\dot y_{k'}y'_{k'})^2( \dot z^2 - 3{y'}^2 )
\nn\\
&&	+ (\dot z^2 +{y'}^2 )(5 {y'}^4
	-2 {y'}^2 \dot z^2 + \dot z^4 )
	+\dot y^2 (4 (\dot y_{k'} y'_{k'})^2 +3{y'}^4
	-2 {y'}^2 \dot z^2 + 3 \dot z^4 ) \Bigr]
\nn\\
&&	+(16 \dot y_{k'} y'_{k'} \dot z_k z'_k +8 (\dot z_k z'_k)^2)((\dot x^A)^2 - 3{y'}^2 )
	-2 {z'}^2 \Bigl[
	\dot y^4 - 6\dot y^2 {y'}^2 -15 {y'}^4 + 2\dot y^2 \dot z^2 
\nn\\
&&	-6 {y'}^2\dot z^2
	+\dot z^4 + 12(\dot x^A {x'}^A)^2 \Bigr]
	+6{z'}^4( (\dot x^A)^2 + 5{y'}^2) 
	+ 10 {z'}^6
	-2p_-^2\Bigl\{
	y^2 \Bigl[
	\dot y^4 - 6\dot y^2 {y'}^2 
\nn\\
&&	-15{y'}^4 -4{y'}^2\dot z^2 -\dot z^4
	+4(\dot x^A {x'}^A)(3\dot y_{k'}y'_{k'} +\dot z_k z'_k )
	-2{z'}^2(2\dot y^2+12 {y'}^2+\dot z^2 ) - 9{z'}^4 \Bigr]
\nn\\
&&	+z^2 \Bigl[
	3\dot y^4 +4(\dot y_{k'} y'_{k'})^2-2\dot y^2 {y'}^2+3{y'}^4
	+4\dot y^2 \dot z^2 +\dot z^4 
	-4(\dot z_k z'_k)^2
\nn\\
&&	+2{z'}^2(6{y'}^2+\dot z^2) 
	+ 9{z'}^4 \Bigr]\Bigr\}
	+p_-^4 \Bigl\{
	y^4\left[
	\dot y^2 + 9{y'}^2 + 2(\dot z^2 + {z'}^2)\right]
	- 8 y^2 z^2 ({x'}^A)^2
\nn\\
&&	+z^4\left[
	2\dot y^2 + 2{y'}^2 +\dot z^2 + 9{z'}^2\right] \Bigr\}
	-p_-^6 (x^A)^2 \left(y^4 + y^2 z^2 + z^4 \right) \biggr\}\ .
\ee
Likewise, the curvature corrections to the worldsheet metric
are found at $O(\Rhat^{-2})$ to be
\be
h_{(2)}^{00} = \frac{1}{2}(z^2-y^2)
	-\frac{1}{2p_-^2}\left[(\dot x^A)^2 + ({x'}^A)^2\right]
\qquad 
h_{(2)}^{01} = \frac{1}{p_-^2}\dot x^A {x'}^A\ .
\ee
Continuing to $O(\Rhat^{-4})$ in the expansion, we find
\be
h_{(4)}^{00} &=& -\frac{1}{8 p_-^4}\biggl\{
	-4 (\dot x^A {x'}^A)^2 
	+ \left[3(\dot x^A)^2 - ({x'}^A)^2\right]
		\left[(\dot x^A)^2 + ({x'}^A)^2\right]
\nn\\
&&	-2p_-^2\Bigl[
	y^2(2{y'}^2-\dot z^2+{z'}^2)
	+z^2(3 \dot y^2+{y'}^2 +2\dot z^2) \Bigr]
	+p_-^4 (y^2 -z^2)^2 \biggr\}\ ,
\nn\\
h_{(4)}^{01} &=& \frac{1}{2p_-^4}\biggl\{
	\dot x^A {x'}^A \left[ (\dot x^A)^2 - ({x'}^A)^2 \right]
	-p_-^2\Bigl[
	y^2 \dot z_k z'_k +\dot y_{k'} y'_{k'} (2y^2 +z^2 ) \Bigr]
	\biggr\}\ . 
\ee
The remaining metric components are fixed by setting $\det h = -1$.

At this stage we have all the information necessary to compute the
Hamiltonian as the generator of time translations on the worldsheet:
\be
H = \frac{\delta {\cal L}}{\delta \dot x^+}\ .
\ee
As a final step we must quantize the theory by converting all
fields to conjugate coordinates and momenta.
The momenta of the transverse $SO(8)$ coordinates
$p_A = {\delta {\cal L}}/{\delta \dot x^A}$
can be computed at each order in the expansion.  The $p_A$ are
then substituted into the Hamiltonian order-by-order, so that all
coordinates $\dot x^A$ are replaced by $p_A$ plus higher-order corrections.
For completeness, we record these corrections for the transverse $SO(4)_{AdS}$
and $SO(4)_{S^5}$ momenta:
\be
(p_z)_k & = & \dot z_k + \frac{1}{2 p_-^2 \Rhat^2}\biggl\{
	-2 z'_k (\dot x^A {x'}^A ) 
	+ \dot z_k ( (\dot x^A)^2 + ({x'}^A)^2 )
	+ p_-^2 \dot z_k y^2 \biggr\}
\nn\\
&&	+\frac{1}{16p_-^4 \Rhat^4}\biggl\{
	8 z'_k \Bigl[
	-(\dot x^A {x'}^A )((\dot x^A)^2 - ({x'}^A)^2 )
	+p_-^2 (2 y^2 \dot y\cdot y' + (y^2-z^2)\dot z\cdot z') \Bigr]
\nn\\
&&	+\dot z_k \biggl[
	2\Bigl[
	-4(\dot x^A {x'}^A )^2 + (3(\dot x^A)^2 - ({x'}^A)^2 )
	((\dot x^A)^2 + ({x'}^A)^2 ) \Bigr]
\nn\\
&&	-4p_-^2\Bigl[
	y^2(2{y'}^2-\dot z^2 +{z'}^2)
	+z^2(2\dot y^2 + \dot z^2 - {z'}^2 )\Bigr]
	+ p_-^4(2 y^4 + z^4 ) \biggr] \biggr\} + O(\Rhat^{-6})\ ,
\nn\\
&&
\ee
\be
(p_y)_{k'} & = & \dot y_{k'} + \frac{1}{2p_-^2 \Rhat^2}\biggl\{
	-2y'_{k'}(\dot x^A {x'}^A) 
	+ \dot y_{k'}((\dot x^A)^2 + ({x'}^A)^2)
	-p_-^2\dot y_{k'}z^2 \biggr\}
\nn\\
&&	+\frac{1}{16p_-^4\Rhat^4}\biggl\{
	2p_-^2\biggl[
	4 y'_{k'}(\dot x^A {x'}^A)(({x'}^A)^2 - (\dot x^A)^2)
	+\dot y_{k'}\Bigl[
	-4(\dot x^A {x'}^A)^2
\nn\\
&&	+(3(\dot x^A)^2-({x'}^A)^2)
	((\dot x^A)^2 + ({x'}^A)^2) \Bigr] \biggr]
	-2 p_-^2 y'_{k'} \Bigl[
	3 y^2 \dot y\cdot y' + 2y^2 \dot z\cdot z' + z^2\dot y\cdot y'
	\Bigr]
\nn\\
&&	+p_-^4\dot y_{k'}(y^4 + 2 z^4) \biggr\} + O(\Rhat^{-6})\ .
\ee

The final result for the Hamiltonian, computed to $O(\Rhat^{-4})$
in the expansion, is
\begin{eqnarray}
{H}^{(0)}_{\rm pp-wave} & = &
    \frac{1}{2}\left[(x^A)^2 + (p_A)^2 + ({x'}^A)^2\right]\ ,
\label{Hpp}
\end{eqnarray}
\begin{eqnarray}
\label{HBBfinal}
{H}^{(2)} & = & 
    \frac{1}{4}\left[ z^2\left( p_{y}^2 + {y'}^2 + 2{z'}^2 \right)
    -y^2\left( p_z^2 + {z'}^2 + 2{y'}^2\right)\right]
    + \frac{1}{8}\left[ (x^A)^2 \right]^2
\nn\\
& &     - \frac{1}{8}\left\{
    \left[ (p_A)^2\right]^2 + 2(p_A)^2({x'}^A)^2
    + \left[ ({x'}^A)^2\right]^2 \right\}
     + \frac{1}{2}\left({x'}^A p_A\right)^2\ ,
\end{eqnarray}
\be
H^{(4)} & = & \frac{1}{32 p_-^5}\biggl\{
	2(p_y^6 + p_z^6) - 8 p_z^2(p_z\cdot z')^2+6p_z^4{y'}^2
	-8{y'}^2(p_z\cdot z')^2 + 6p_z^2{y'}^4 + 2{y'}^6 + 6p_-^2p_z^4y^2
\nn\\
&&	-24 p_-^2y^2(p_z\cdot z')^2 + 16p_-^2y^2 {y'}^2p_z^2
	+10p_-^2y^2{y'}^4+2p_-^4p_z^2 y^4 + 9p_-^4{y'}^2 y^4
	+p_-^6y^6 
\nn\\
&&	+ 6p_z^4{z'}^2 - 8{z'}^2(p_z\cdot z')^2 
	+ 12 p_z^2 {y'}^2{z'}^2 + 6{y'}^4{z'}^2
	+ 12p_-^2 p_z^2 y^2 {z'}^2 
\nn\\
&&	+ 16 p_-^2 {y'}^2 y^2 {z'}^2
	+2p_-^4 y^4{z'}^2 + 6p_z^2{z'}^4 + 6{y'}^2{z'}^4
	+6p_-^2 y^2{z'}^4 + 2{z'}^6
	+2p_y^4\Bigl[
	3p_z^2+3({x'}^A)^2
\nn\\
&&	+p_-^2(y^2-z^2) \Bigr]
	-\Bigl[8(p_y\cdot y')^2 +  16(p_y\cdot y')(p_z\cdot z')\Bigr]\Bigl[
	p_z^2+({x'}^A)^2+p_-^2(3y^2-z^2)\Bigr]
\nn\\
&&	+2p_-^2 z^2 \Bigl[
	p_z^4 + 4(p_z\cdot z')^2 - {y'}^4-4p_-^2{y'}^2y^2+p_-^4y^4
	-2{z'}^2(p_z^2+2{y'}^2+2p_-^2y^2)-3{z'}^4\Bigr]
\nn\\
&&	+p_-^4 z^4(2{y'}^2-p_z^2+2p_-^2y^2 + 9{z'}^2)
	+p_-^6 z^6
	+p_y^2\Bigl[
	6p_z^4 - 8(p_A {x'}^A)^2 + 6{y'}^4 + 12p_-2{y'}^2y^2
\nn\\
&&	-p_-^4y^4+12{y'}^2{z'}^2 + 8p_-^2y^2{z'}^2 +6{z'}^4
	+4p_z^2(3{y'}^2+2p_-^2 y^2 + 3{z'}^2 )
\nn\\
&&	-4p_-^2 z^2({y'}^2+2 {z'}^2)
	+2p_-^4 z^4 \Bigr] \biggr\}\ .
\ee
As expected, the leading-order system is exactly the quadratic 
pp-wave Hamiltonian originally reported by Metsaev in 
\cite{Metsaev:2001bj}.  Similarly, $H^{(2)}$ agrees with the bosonic 
sector of the $O(1/\Rhat^2)$ quartic Hamiltonian computed in 
\cite{Callan:2003xr,Callan:2004uv}.  

Since we are only interested in the closed bosonic
subsectors that are restricted to each of the $SO(4)_{AdS}$ and
$SO(4)_{S^5}$ subspaces, the results at $O(1/\Rhat^{4})$ can be 
dramatically simplified by projecting $H^{(4)}$ onto these subspaces:
\be
H_{AdS}^{(4)} & = & \frac{1}{32p_-^5}\biggl\{
	2(p_z^2+{z'}^2)(p_z^2-2p_z\cdot z' + {z'}^2)(p_z^2+2p_z\cdot z' + {z'}^2)
\nn\\
&&\kern-30pt	+2p_-^2 z^2\Bigl[
	4(p_z\cdot z')^2 + (p_z^2-3{z'}^2)(p_z^2+{z'}^2)\Bigr]
	-p_-^4z^4(p_z^2-9{z'}^2) + p_-^6 z^6 \biggr\}\ ,
\label{HFSO41}
\ee
\be
H_{S^5}^{(4)} & = & \frac{1}{32p_-^5}\biggl\{
	2(p_y^2+{y'}^2)(p_y^2-2p_y\cdot y' + {y'}^2)(p_y^2+2p_y\cdot y' + {y'}^2)
\nn\\
&&\kern-30pt	+2p_-^2 y^2\Bigl[
	-12(p_y\cdot y')^2 + (p_y^2+{y'}^2)(p_y^2+5{y'}^2)\Bigr]
	-p_-^4y^4(p_y^2-9{y'}^2) + p_-^6 y^6 \biggr\}\ .
\label{HFSO42}
\ee
At this point the Hamiltonian can be expanded in terms of
raising and lowering operators, and matrix elements 
can be computed between states lying in the closed bosonic subsectors
of the theory.

\section{Matrix elements and integrability}

The unperturbed string eigenstates are the exact eigenstates
of the pp-wave background, and the ground state $\ket{J}$ 
(of the bosonic theory) is that of the eight transverse 
string oscillators satisfying
\be
\ddot x^A - {x''}^A + p_-^2 x^A = 0\ ,
\ee
and carrying angular momentum $J$ on the $S^5$ subspace.
This is solved by the usual Fourier expansion
\be
x^A(\sigma,\tau) & = & \sum_{n=-\infty}^\infty x_n^A(\tau)e^{-ik_n\sigma}\ ,\nn\\
x_n^A(\tau) & = & \frac{i}{\sqrt{2\omega_n}}\left(
		a_n^A e^{-i\omega_n\tau}-a_{-n}^{\dag A}e^{i\omega_n\tau}\right)\ ,
\ee
with integer mode numbers $k_n = n$ and $\omega_n$ defined by 
$\omega_n \equiv \sqrt{p_-^2 + k_n^2}$.
The creation and annihilation operators obey the standard relation
$[ a_m^A,a_n^{\dag B}] = \delta_{mn}\delta^{AB}$, in terms of which the 
pp-wave Hamiltonian (\ref{Hpp}) takes the form
\be
H_{\rm pp-wave}^{(0)} = \frac{1}{p_-}\sum_{n=-\infty}^\infty\omega_n
		\left( a_n^{\dag A} a_n^A +4 \right)\ .
\ee
The zero-point term is canceled when fermions are included.

As noted above, the string theory version of the SYM parity operator,
which inverts the ordering of fields within single-trace operators,
is one that invokes an overall sign change on the worldsheet mode
indices of the unperturbed eigenstates:
\be
P (a_{q_1}^{A_1\dag} a_{q_2}^{A_2\dag} \dots )\ket{J} =  
	a_{-q_1}^{A_1\dag} a_{-q_2}^{A_2\dag}  \dots \ket{J}\ .
\ee
In the language of spin chains, $P$ acts 
in a similar fashion on pseudoparticle states
by applying a sign switch to the lattice momenta of each state.
At least three worldsheet (spin-chain) impurities 
with non-zero momenta are required to admit string
(pseudoparticle) states that are distinct under the action of $P$. 

As described in 
\cite{3IMP}, the bosonic three-impurity string Fock space consists
of the $512$-dimensional space spanned by the states
\be
a_q^{\dag A} a_r^{\dag B} a_s^{\dag C} \Ket{J}\ ,
\nn
\ee
subject to the level-matching constraint $q+r+s = 0$. 
The upper-case indices $A,B,C,\dots = 1,\dots,8$ span the transverse $SO(8)$,
and the lower-case notation $a,b,c = 1,\dots,4$ and $a',b',c' = 5,\dots,8$ 
will be used to indicate vectors in the $SO(4)_{AdS}$ and $SO(4)_{S^5}$
subspaces, respectively.  
Within the bosonic sector of the theory there are two 
subsectors which decouple at all orders in $\lambda'$.  
These subsectors consist of symmetrized, traceless bosonic impurities 
restricted to lie in either $SO(4)_{AdS}$ or $SO(4)_{S^5}$:
\be
a_q^{\dag (a} a_r^{\dag b} a_s^{\dag c)} \ket{J}\ , 
 \qquad
a_q^{\dag (a'} a_r^{\dag b'} a_s^{\dag c')} \ket{J}\ .\nn
\ee
(Here, tracelessness implies $a\neq b\neq c$ and $a'\neq b'\neq c'$.)
By restricting to these protected subsectors, we can compute
matrix elements that do not mix with any other sectors of the theory
and are exact in $\lambda'$.  The three-impurity block-diagonalization
of the Hamiltonian on these closed bosonic subsectors was demonstrated 
in more detail in \cite{3IMP}.

To $O(\Rhat^{-2})$, the matrix elements in these sectors
were reported in \cite{3IMP}.  At this order, the Hamiltonian is 
quartic in oscillators and matrix elements taken between three-impurity
string states always involve a single forward-scattering contraction.
This fact breaks the state space into two distinct classes, where 
the mode indices $(q,r,s)$ are either completely inequivalent
$(q\neq r\neq s)$, or two mode indices are taken to be equal
($q=r=n$, $s=-2n$).  At $O(\Rhat^{-6})$, this complication does not
arise:  the Hamiltonian is sixth-order in fields and matrix elements taken 
in the three-impurity regime do not involve
contractions taken directly between the external unperturbed 
eigenstates. 

Using the AdS/CFT relation
\be
\Rhat^2 \rightleftharpoons p_- J\ ,
\ee
all instances of the radius $\Rhat$
will henceforth be replaced with the $S^5$ angular momentum $J$.  
In the $SO(4)_{AdS}$ subspace, to $O(1/J^2)$, the Hamiltonian in the 
closed symmetric-traceless subsector 
exhibits the following matrix elements
\be
\Bra{J } &&\kern-20pt a_q^{( a}a_r^{b}a_s^{c )} ( H^{(4)}_{AdS} )
	a_s^{\dag (a}a_r^{\dag b}a_q^{\dag c )} \Ket{J}  = 
	\frac{1}{4 J^2 {\lambda'}^{3/2}\omega_q\omega_r\omega_s}\biggl\{
	15+\lambda'\Bigl[
	9rs+16s^2
\nn\\
&&	-3q(r+s)(-3+2rs\lambda')
	+2r^2(8+5s^2\lambda')+2q^2(8-3rs\lambda'+5s^2\lambda'
\nn\\
&&	+r^2\lambda'(5+12s^2\lambda'))\Bigr]
	+\lambda'\omega_r\omega_s
	\Bigl[1+2\lambda'(-rs+q(r+s)+q^2(1-4rs\lambda'))\Bigr]
\nn\\
&&	+\lambda'\omega_q\biggl[
	(1+2s(r+s)\lambda'-2q\lambda'(r-s+4rs^2\lambda'))\omega_r
	+\omega_s\Bigl[1+2r(r+s)\lambda'
\nn\\
&&\kern+0pt	-2q\lambda'(s+r(-1+4rs\lambda'))
	\Bigr]\biggr]\biggr\}\ .
\label{AdSME}
\ee
The generic upper indices $a,b,c$ are taken to be any indices in the $SO(4)_{AdS}$
subspace ($a,b,c = 1,\dots,4$) and, because the state is traceless, $a\neq b\neq c$. 
We can perform a small-$\lambda'$ expansion to obtain
\be
\Bra{J } &&\kern-20pt a_q^{( a}a_r^{b}a_s^{c )} ( H^{(4)}_{AdS} )
	a_s^{\dag (a}a_r^{\dag b}a_q^{\dag c )} \Ket{J} 
 = \frac{9}{2J^2} + \frac{9\lambda'}{4J^2}(q^2+qr+r^2)
	+\frac{11{\lambda'}^2}{16J^2}(q^2+qr+r^2)^2
\nn\\
&&\kern-20pt	-\frac{3{\lambda'}^3}{64J^2}\biggl[
	14q^6+42q^5r-51q^4r^2-172q^3r^3-51q^2r^4+42qr^5+14r^6\biggr] + O({\lambda'}^4)\ .
\label{AdSMEExp}
\ee
Here we have made the substitution $s=-q-r$ (using the level-matching constraint) 
to simplify the resulting expression.
In the $SO(4)_{S^5}$ subspace, we find
\be
\Bra{J } &&\kern-20pt a_q^{( a'}a_r^{b'}a_s^{c' )} ( H^{(4)}_{S^5} )
	a_s^{\dag (a'}a_r^{\dag b'}a_q^{\dag c' )} \Ket{J}  = 
	\frac{1}{4 J^2 {\lambda'}^{3/2}\omega_q\omega_r\omega_s}\biggl\{
	15+\lambda'\Bigl[
	9rs + 16s^2 
\nn\\
&&	+q(r+s)(9+10rs\lambda')
	+2r^2(8+9s^2\lambda')+2q^2(8+5rs\lambda' +9s^2\lambda'
	+3r^2\lambda'(3+4s^2\lambda')) \Bigr]
\nn\\
&&	-\omega_q\omega_r\Bigl[
	-1+6rs\lambda' -2s^2\lambda' + 2q\lambda'(3s+r(5+4s^2\lambda'))\Bigr]
	-\omega_s\lambda'\biggl[\omega_q \Bigl[
	-1-2r(r-3s)\lambda'
\nn\\
&&	+2q\lambda'(5s+r(3+4rs))\Bigr]
	+\omega_r\Bigl[-1+2\lambda'(5rs+3q(r+s)+q^2(-1+4rs\lambda'))\Bigr]
	\biggr]\biggr\}
\nn\\
&&\kern-20pt	{= }~ \frac{9}{2J^2} + \frac{33}{4J^2}(q^2+qr+r^2)\lambda'
	+\frac{11}{16J^2}(q^2+qr+r^2)^2{\lambda'}^2
\nn\\
&&\kern-20pt	-\frac{{\lambda'}^3}{64J^2}\biggl[
	74q^6+222q^5r + 39q^4 r^2 - 292q^3r^3
	+39q^2 r^4 + 222q r^5 + 74 r^6 \biggr] +O({\lambda'}^4)\ .
\label{S5ME}
\ee
Again, we have expanded in small $\lambda'$, setting $s=-q-r$ in the end.
The indices $a',b',c' = 5,\dots,8$ lie in $SO(4)_{S^5}$, and are again
taken to be inequivalent.  

To find the full energy shifts of these states
to $O(1/J^2)$, we would need to compute the full contribution from $H^{(2)}$
in second-order perturbation theory.  As noted above, such calculations
require knowledge of how the theory is renormalized at this order, and
these issues will be dealt with elsewhere.  We can avoid these complications
in the present setting by restricting to zeroth order 
in the small-$\lambda'$ expansion:  the salient points regarding integrability
can still be made at this level.  

To $O({\lambda'}^0)$, all energy levels
are degenerate, and a trivial consequence of this fact is that,
to any order in $1/J$,
\be
\left[ H_{AdS},P \right] = 0 + O(\lambda')\ , \qquad 
\left[ H_{S^5},P \right] = 0 + O(\lambda')\ .
\label{paritycheck}
\ee
To test integrability, we aim to determine whether the Hamiltonian also commutes
with $Q_2^{\rm string}(\lambda')$ to this order. Since
$Q_2^{\rm string}(\lambda')$ must anticommute with $P$ and 
connect degenerate parity pairs (by definition), and because 
the Hamiltonian commutes with $P$ to $O({\lambda'}^0)$, 
$Q_2^{\rm string}(\lambda')$
can only commute with the Hamiltonian if the Hamiltonian itself
does not connect states of opposite parity at zeroth order
in $\lambda'$.  To test whether $Q_2^{\rm string}(\lambda')$ is truly 
a conserved charge in the theory, we can therefore compute matrix elements 
that connect string states of opposite parity.   
Starting with the first-order contributions from $H^{(4)}$ on the $SO(4)_{AdS}$ side,
we find
\be
\Bra{J } &&\kern-20pt a_{-q}^{( a}a_{-r}^{b\phantom{)}}a_{-s}^{c )} ( H^{(4)}_{AdS} )
	a_s^{\dag (a}a_r^{\dag b}a_q^{\dag c )} \Ket{J}  = 
	\frac{1}{4 J^2 {\lambda'}^{3/2}\omega_q\omega_r\omega_s}\biggl\{
	15 + \lambda'\Bigl[
	-9rs+5s^2
\nn\\
&&	-3q(r+s)(3+2rs\lambda')
	+r^2(5+6s^2\lambda')+q^2(5+6(r^2-rs+s^2)\lambda')\Bigr]
\nn\\
&&	+\lambda'\omega_r\omega_s\Bigl[
	1+2(q^2-3rs+q(r+s))\lambda'\Bigr]
	+\lambda'\omega_q\Bigl[
	\omega_r(1+2q(s-3r)\lambda'+2s(r+s)\lambda')
\nn\\
&&	+\omega_s(1+2q(r-3s)\lambda'+2r(r+s)\lambda')\Bigr]
	\biggr\}
\nn\\
&&\kern-20pt =~ \frac{9}{2J^2} + \frac{9}{4J^2}(q^2+qr+r^2)\lambda'
	+\frac{11}{16J^2}(q^2+qr+r^2)^2{\lambda'}^2
\nn\\
&&\kern-20pt	-\frac{3{\lambda'}^3}{64J^2}\biggl[
	14q^6+42q^5r+165q^4r^2+260q^3r^3+165q^2r^4+42qr^5+14r^6\biggr]
	 + O({\lambda'}^4)\ .
\nn\\
&&
\label{AdSMX}
\ee
In the $SO(4)_{S^5}$ subspace, we have
\be
\Bra{J } &&\kern-20pt a_{-q}^{( a'}a_{-r}^{b'\phantom{)}}a_{-s}^{c' )} ( H^{(4)}_{S^5} )
	a_s^{\dag (a'}a_r^{\dag b'}a_q^{\dag c' )} \Ket{J}  = 
	\frac{1}{4 J^2 {\lambda'}^{3/2}\omega_q\omega_r\omega_s}\biggl\{
	15 +\lambda'\Bigl[
	-9rs + 5s^2 
\nn\\
&&	+q(r+s)(-9+10rs\lambda')
	+r^2(5+14s^2\lambda')+q^2(5+2(7r^2+5rs+7s^2)\lambda')\Bigr]
\nn\\
&&	+\lambda'\omega_r\omega_s\Bigl[
	1+2(q^2-7rs-3q(r+s))\lambda'\Bigr]
	+\lambda'\omega_q\biggl[
	(1+2(-7qr-3(q+r)s+s^2)\lambda')\omega_r
\nn\\
&&	+\Bigl[1+2r(r-3s)\lambda'-2q(3r+7s)\lambda'\Bigr]
	\omega_s\biggr]
	\biggr\}
\nn\\
&&\kern-20pt {= }~
	\frac{9}{2J^2}+\frac{33}{4J^2}(q^2+qr+r^2)\lambda'
	+\frac{11}{16J^2}(q^2+qr+r^2)^2{\lambda'}^2
\nn\\
&&\kern-20pt	-\frac{{\lambda'}^3}{64J^2}\biggl[
	74q^6+222q^5r+687q^4r^2+1004 q^3r^3
	+687q^2r^4+222qr^5+74r^6\biggr]
	+O({\lambda'}^4)\ .
\nn\\
&&
\label{S5MX}
\ee

The Hamiltonian is {\it a priori} $2\times 2$ block-diagonal in this 
three-impurity basis of degenerate parity pairs.  The theory, however,
does not conserve impurity number, and the unperturbed eigenstates 
above do not constitute a complete basis.  To properly compute these 
matrix elements to the order of interest, one must include higher 
perturbative corrections to the zeroth-order three-impurity basis 
states that involve different numbers of excitations in the intermediate 
channels.  Denoting, for example, the zeroth-order $SO(4)_{S^5}$ 
eigenstates above as
\be
\Ket{+^{(0)}} = a_s^{\dag (a'}a_r^{\dag b'}a_q^{\dag c' )} \Ket{J}\ 
\qquad 
\Ket{-^{(0)}} = a_{-s}^{\dag (a'}a_{-r}^{\dag b'}a_{-q}^{\dag c' )} \Ket{J}\ ,
\ee
the full matrix element in eqns.~(\ref{AdSMX},\ref{S5MX}), to $O(1/J^2)$, takes the form
\be
\braket{-^{(0)}|H|+^{(0)}} & = & 
	\braket{-^{(0)}|H^{(4)}|+^{(0)}} 
	+ \sum_{\psi \neq \pm } 
	\frac{\braket{-^{(0)}|H^{(2)}|\psi^{(0)}}
	\braket{\psi^{(0)}|H^{(2)}|+^{(0)}} }{E_\pm^{(0)} - E_\psi^{(0)}  }\ . 
\label{2ndorder}
\ee
The intermediate state sum, mediated by $H^{(2)}$, can in principle involve 
transitions that mix the three-impurity states $\Ket{+^{(0)}}$ 
and $\Ket{-^{(0)}}$ with any one-, three-, five- and seven-impurity states
in the theory.  
In addition to the 
purely bosonic sector $H^{(2)}_{\rm BB}$, these sums can also involve the 
pure-fermi sector $H^{(2)}_{\rm FF}$ and
the bose-fermi mixing sector $H^{(2)}_{\rm BF}$ (see \cite{Callan:2003xr,Callan:2004uv}
for details).  Some of these summations are excluded by simple arguments, 
however.  

Starting with the pure-boson sector $H^{(2)}_{\rm BB}$, intermediate
channels involving one excitation vanish by direct calculation
in both the $SO(4)_{AdS}$ and $SO(4)_{S^5}$ subspaces:
\be
\braket{J| a_{p_1}^A (H^{(2)}_{\rm BB}) a_s^{\dag (a}a_r^{\dag b}a_q^{\dag c )} |J}
	= 0
\qquad
\braket{J| a_{p_1}^A (H^{(2)}_{\rm BB}) a_s^{\dag (a'}a_r^{\dag b'}a_q^{\dag c' )} |J}
	= 0\ .
\ee
The three-to-three impurity channel has no contributions at $O({\lambda'}^0)$:
\be
&&\kern-20pt 
\braket{J| a_{p_1}^{A_1} a_{p_2}^{A_2} a_{p_3}^{A_3} 
	(H^{(2)}_{\rm BB}) a_s^{\dag (a}a_r^{\dag b}a_q^{\dag c )} |J}
	= O(\lambda')
\qquad
\nn\\
&&\kern-25pt 
\braket{J| a_{p_1}^{A_1} a_{p_2}^{A_2} a_{p_3}^{A_3}  
	(H^{(2)}_{\rm BB}) a_s^{\dag (a'}a_r^{\dag b'}a_q^{\dag c' )} |J}
	= O(\lambda')\ .
\label{315}
\ee
It should be noted that, since the propagator in this channel is nonzero at 
$O({\lambda'}^{-1})$, contributions from eqn.~(\ref{315}) at $O({\lambda'}^{1/2})$ 
could potentially affect the final result.
The matrix elements in (\ref{315}) vanish to $O(\lambda')$, however, and
this is not a concern.
The zeroth-order three-to-five impurity contributions to eqn.~(\ref{2ndorder}) 
from the $SO(4)_{AdS}$ and $SO(4)_{S^5}$ sectors are
\be
-\frac{1}{2} \sum_{p_i, A_i} &&\kern-20pt
	\biggl[
	\Bra{J}
	a_{-q}^{(a}a_{-r}^{b}a_{-s}^{c )} 
	(H_{\rm BB}^{(2)})
	a_{p_1}^{\dag A_1}a_{p_2}^{\dag A_2}a_{p_3}^{\dag A_3}
	a_{p_4}^{\dag A_4}a_{p_5}^{\dag A_5}  \Ket{J}
\nn\\
&&\kern-0pt	\times \Bra{J}
	a_{p_5}^{A_5}a_{p_4}^{A_4}
	a_{p_3}^{A_3}a_{p_2}^{A_2}
	a_{p_1}^{A_1}
	(H_{\rm BB}^{(2)})
	a_s^{(\dag a}a_r^{\dag b}a_q^{\dag c )} 
	\Ket{J}
	\biggr]
	= -\frac{675}{J^2} + O({\lambda'})
\nn\\
-\frac{1}{2} \sum_{p_i, A_i} &&\kern-20pt
	\biggl[
	\Bra{J}
	a_{-q}^{(a'}a_{-r}^{b'}a_{-s}^{c' )} 
	(H_{\rm BB}^{(2)})
	a_{p_1}^{\dag A_1}a_{p_2}^{\dag A_2}a_{p_3}^{\dag A_3}
	a_{p_4}^{\dag A_4}a_{p_5}^{\dag A_5}  \Ket{J}
\nn\\
&&\kern-0pt	\times \Bra{J}
	a_{p_5}^{A_5}a_{p_4}^{A_4}
	a_{p_3}^{A_3}a_{p_2}^{A_2}
	a_{p_1}^{A_1}
	(H_{\rm BB}^{(2)})
	a_s^{(\dag a'}a_r^{\dag b'}a_q^{\dag c' )} 
	\Ket{J}
	\biggr]
	= -\frac{1,755}{J^2} + O({\lambda'})\ .
\ee
Finally, three-to-seven impurity contributions from $H^{(2)}_{\rm BB}$ vanish to this order:
\be
- \frac{1}{4}\sum_{p_i, A_i} &&\kern-20pt
	\biggl[
	\Bra{J}
	a_{-q}^{(a}a_{-r}^{b}a_{-s}^{c )} 
	(H_{\rm BB}^{(2)})
	a_{p_1}^{\dag A_1}a_{p_2}^{\dag A_2}a_{p_3}^{\dag A_3}
	a_{p_4}^{\dag A_4}a_{p_5}^{\dag A_5}a_{p_6}^{\dag A_6}
	a_{p_7}^{\dag A_7}  
	\Ket{J}  
\nn\\
&&\kern+15pt	\times \Bra{J}
	a_{p_7}^{A_7}a_{p_6}^{A_6}a_{p_5}^{A_5}
	a_{p_4}^{A_4}a_{p_3}^{A_3}a_{p_2}^{A_2}a_{p_1}^{A_1}
	(H_{\rm BB}^{(2)})
	a_{s}^{\dag (a}a_{r}^{\dag b}a_{q}^{\dag c )} 
	\Ket{J}
	\biggr]
	= O({\lambda'})\
\nn\\
- \frac{1}{4}\sum_{p_i, A_i} &&\kern-20pt
	\biggl[
	\Bra{J}
	a_{-q}^{(a'}a_{-r}^{b'}a_{-s}^{c' )} 
	(H_{\rm BB}^{(2)})
	a_{p_1}^{\dag A_1}a_{p_2}^{\dag A_2}a_{p_3}^{\dag A_3}
	a_{p_4}^{\dag A_4}a_{p_5}^{\dag A_5}a_{p_6}^{\dag A_6}
	a_{p_7}^{\dag A_7}  
	\Ket{J}  
\nn\\
&&\kern+15pt	\times \Bra{J}
	a_{p_7}^{A_7}a_{p_6}^{A_6}a_{p_5}^{A_5}
	a_{p_4}^{A_4}a_{p_3}^{A_3}a_{p_2}^{A_2}a_{p_1}^{A_1}
	(H_{\rm BB}^{(2)})
	a_{s}^{\dag (a'}a_{r}^{\dag b'}a_{q}^{\dag c' )} 
	\Ket{J}
	\biggr]
	=  O({\lambda'})\ .
\ee

By inspection, the bose-fermi sector $H^{(2)}_{\rm BF}$ cannot mediate 
mixing between three- and one-impurity string states:
\be
&&\kern-20pt
\Bra{J} a_q^{(a}a_r^{b}a_s^{c )} 
	(H_{\rm BF}^{(2)}) a_0^{\dag A} \Ket{J} = 
\Bra{J} a_q^{(a}a_r^{b }a_s^{c  )} 
	(H_{\rm BF}^{(2)}) b_0^{\dag \alpha} \Ket{J} = 0\
\nn\\
&&\kern-28pt
\Bra{J} a_q^{(a'}a_r^{b'}a_s^{c' )} 
	(H_{\rm BF}^{(2)}) a_0^{\dag A} \Ket{J} = 
\Bra{J} a_q^{(a'}a_r^{b'}a_s^{c' )} 
	(H_{\rm BF}^{(2)}) b_0^{\dag \alpha} \Ket{J} = 0\ .
\ee
This sector can mix three bosonic impurities with spacetime
boson states comprised of a single bosonic excitation and two
fermionic excitations.  At $O({\lambda'}^0)$, however,
there are no three-to-three impurity matrix elements of $H^{(2)}_{\rm BF}$
in this channel:
\be
&&\kern-20pt	\Bra{J}
	a_q^{(a}a_r^{b}a_s^{c )} 
	(H_{\rm BF}^{(2)})
	b_{p_1}^{\dag \alpha_1}b_{p_2}^{\dag \alpha_2}a_{p_3}^{\dag A} \Ket{J} 
	= O({\lambda'})
\nn\\
&&\kern-25pt	\Bra{J}
	a_q^{(a'}a_r^{b'}a_s^{c' )} 
	(H_{\rm BF}^{(2)})
	b_{p_1}^{\dag \alpha_1}b_{p_2}^{\dag \alpha_2}a_{p_3}^{\dag A} \Ket{J} 
	= O({\lambda'})\ .
\ee
The three-to-five impurity interaction gives equal contributions from
the $AdS_5$ and $S^5$ sides:
\be
-\frac{1}{2} \sum_{p_i, A_i, \alpha_i} &&\kern-20pt
	\biggl[
	\Bra{J}
	a_{-q}^{(a}a_{-r}^{b}a_{-s}^{c )} 
	(H_{\rm BF}^{(2)})
	b_{p_1}^{\dag \alpha_1}b_{p_2}^{\dag \alpha_2}a_{p_3}^{\dag A_1}
	a_{p_4}^{\dag A_2}a_{p_5}^{\dag A_3}  \Ket{J}
\nn\\
&&\kern-0pt	\times \Bra{J}
	a_{p_5}^{ A_3}a_{p_4}^{ A_2} a_{p_3}^{ A_1}
	b_{p_2}^{ \alpha_2}b_{p_1}^{ \alpha_1}
	(H_{\rm BF}^{(2)})
	a_s^{\dag (a}a_r^{\dag b}a_q^{\dag c )}
	\Ket{J}
	\biggr]
	= -\frac{10}{9J^2} + O({\lambda'})\ 
\nn\\
-\frac{1}{2} \sum_{p_i, A_i, \alpha_i} &&\kern-20pt
	\biggl[
	\Bra{J}
	a_{-q}^{(a'}a_{-r}^{b'}a_{-s}^{c' )} 
	(H_{\rm BF}^{(2)})
	b_{p_1}^{\dag \alpha_1}b_{p_2}^{\dag \alpha_2}a_{p_3}^{\dag A_1}
	a_{p_4}^{\dag A_2}a_{p_5}^{\dag A_3}  \Ket{J}
\nn\\
&&\kern-0pt	\times \Bra{J}
	a_{p_5}^{ A_3}a_{p_4}^{ A_2} a_{p_3}^{ A_1}
	b_{p_2}^{ \alpha_2}b_{p_1}^{ \alpha_1}
	(H_{\rm BF}^{(2)})
	a_s^{\dag (a'}a_r^{\dag b'}a_q^{\dag c' )}
	\Ket{J}
	\biggr]
	= -\frac{10}{9J^2} + O({\lambda'})\ .
\ee
The three-to-seven $H_{\rm BF}^{(2)}$ channel yields
\be
-\frac{1}{4} \sum_{p_i, A_i, \alpha_i} &&\kern-20pt
	\biggl[
	\Bra{J}
	a_{-q}^{(a}a_{-r}^{b}a_{-s}^{c )} 
	(H_{\rm BF}^{(2)})
	b_{p_1}^{\dag \alpha_1}b_{p_2}^{\dag \alpha_2}a_{p_3}^{\dag A_1}
	a_{p_4}^{\dag A_2}a_{p_5}^{\dag A_3}a_{p_6}^{\dag A_4}a_{p_7}^{\dag A_5}  
	\Ket{J} 
\nn\\
&&\kern+15pt	\times \Bra{J}
	a_{p_7}^{A_5}a_{p_6}^{ A_4}a_{p_5}^{ A_3}a_{p_4}^{ A_2}
	a_{p_3}^{ A_1}b_{p_2}^{\alpha_2}b_{p_1}^{ \alpha_1}
	(H_{\rm BF}^{(2)})
	a_s^{\dag (a}a_r^{\dag b}a_q^{\dag c )}
	\Ket{J}
	\biggr]
	 = O(\lambda')\ 
\nn\\
-\frac{1}{4} \sum_{p_i, A_i, \alpha_i} &&\kern-20pt
	\biggl[
	\Bra{J}
	a_{-q}^{(a'}a_{-r}^{b'}a_{-s}^{c' )} 
	(H_{\rm BF}^{(2)})
	b_{p_1}^{\dag \alpha_1}b_{p_2}^{\dag \alpha_2}a_{p_3}^{\dag A_1}
	a_{p_4}^{\dag A_2}a_{p_5}^{\dag A_3}a_{p_6}^{\dag A_4}a_{p_7}^{\dag A_5}  
	\Ket{J} 
\nn\\
&&\kern+15pt	\times \Bra{J}
	a_{p_7}^{A_5}a_{p_6}^{ A_4}a_{p_5}^{ A_3}a_{p_4}^{ A_2}
	a_{p_3}^{ A_1}b_{p_2}^{\alpha_2}b_{p_1}^{ \alpha_1}
	(H_{\rm BF}^{(2)})
	a_s^{\dag (a'}a_r^{\dag b'}a_q^{\dag c' )}
	\Ket{J}
	\biggr]
	 = O(\lambda')\ .
\ee

The only interaction permitted in the pure-fermi sector $H^{(2)}_{\rm FF}$ 
is the three-to-seven impurity transition with intermediate states composed
of three bosonic excitations and four fermionic excitations.  These contributions
vanish in both the $SO(4)_{AdS}$ and $SO(4)_{S^5}$ sectors:
\be
-\frac{1}{4} \sum_{p_i, A_i, \alpha_i} &&\kern-20pt
	\biggl[
	\Bra{J}
	a_{-q}^{(a}a_{-r}^{b}a_{-s}^{c)} 
	(H_{\rm FF}^{(2)})
	b_{p_1}^{\dag \alpha_1}b_{p_2}^{\dag \alpha_2}
	b_{p_3}^{\dag \alpha_2}b_{p_4}^{\dag \alpha_4}
	a_{p_5}^{\dag A_1}
	a_{p_6}^{\dag A_2}
	a_{p_7}^{\dag A_3}
	\Ket{J} 
\nn\\
&&\kern+15pt	\times \Bra{J}
	a_{p_7}^{ A_3}
	a_{p_6}^{ A_2} 
	a_{p_5}^{ A_1}
	b_{p_4}^{ \alpha_4}b_{p_3}^{ \alpha_2}
	b_{p_2}^{ \alpha_2}b_{p_1}^{ \alpha_1}
	(H_{\rm FF}^{(2)})
	a_{s}^{\dag (a}a_{r}^{\dag b}a_{q}^{\dag c )} 
	\Ket{J}
	\biggr]
	= O(\lambda')\ 
\nn\\
-\frac{1}{4} \sum_{p_i, A_i, \alpha_i} &&\kern-20pt
	\biggl[
	\Bra{J}
	a_{-q}^{(a'}a_{-r}^{b'}a_{-s}^{c' )} 
	(H_{\rm FF}^{(2)})
	b_{p_1}^{\dag \alpha_1}b_{p_2}^{\dag \alpha_2}
	b_{p_3}^{\dag \alpha_2}b_{p_4}^{\dag \alpha_4}
	a_{p_5}^{\dag A_1}
	a_{p_6}^{\dag A_2}
	a_{p_7}^{\dag A_3}
	\Ket{J} 
\nn\\
&&\kern+15pt	\times \Bra{J}
	a_{p_7}^{ A_3}
	a_{p_6}^{ A_2} 
	a_{p_5}^{ A_1}
	b_{p_4}^{ \alpha_4}b_{p_3}^{ \alpha_2}
	b_{p_2}^{ \alpha_2}b_{p_1}^{ \alpha_1}
	(H_{\rm FF}^{(2)})
	a_{s}^{\dag (a'}a_{r}^{\dag b'}a_{q}^{\dag c' )} 
	\Ket{J}
	\biggr]
	= O(\lambda')\ .
\ee
No other contributions can arise from $H^{(2)}_{\rm FF}$ 
(apart from those involving normal-ordering terms, 
which are excluded by supersymmetry \cite{Callan:2004uv}). 

In the end we find that the matrix elements mixing bosonic parity pairs
are nonzero at $O(1/J^2)$:
\be
\Bra{J } 
&&\kern-20pt a_{-q}^{( a'}a_{-r}^{b'\phantom{)}}a_{-s}^{c' )} 
	( H_{S^5} )
	a_s^{\dag (a'}a_r^{\dag b'}a_q^{\dag c' )} \Ket{J}  \neq 0
\nn\\
\Bra{J } &&\kern-20pt 
a_{-q}^{( a}a_{-r}^{b\phantom{)}}a_{-s}^{c )} 
	( H_{AdS} )
	a_s^{\dag (a}a_r^{\dag b}a_q^{\dag c )} \Ket{J}  \neq 0\ .
\label{AdSMX2}
\ee
This is the first order in the $1/J$ expansion where this sort of mixing 
can possibly be observed: at lower orders the Hamiltonian is either quartic 
or quadratic in fields and therefore cannot mix distinct three-impurity states 
connected by parity.

To interpret this mixing in terms of the comparison
with gauge theory dynamics, we first note that operators of definite and distinct parity
cannot mix in ${\cal N}=4$ SYM \cite{D'Hoker:2003vf}.  When string eigenstates are
arranged into states of definite parity there are no off-diagonal matrix elements
of the Hamiltonian that connect states of opposite parity.  This aspect of the 
string theory is therefore in agreement with gauge theory predictions.\footnote{I
thank N.~Beisert for clarification on this point.}
We have also established that to $O(1/J^2)$ and $O({\lambda'}^0)$ 
the string Hamiltonian commutes with the parity operator $P$ (\ref{paritycheck}).  
As noted above, since $Q_2^{\rm string}(\lambda')$ must anticommute with $P$ and 
connect degenerate parity pairs,
it can only commute with the Hamiltonian if the Hamiltonian itself
does not connect states that are themselves connected by parity
(in other words, string eigenstates of definite parity are non-degenerate).  
The results in eqn.~(\ref{AdSMX2}) show that 
$Q_2^{\rm string}(\lambda')$ therefore 
fails to commute with the Hamiltonian at this order:    
\be
\left[ H_{AdS} , Q_2^{\rm string}(\lambda') \right] \neq 0\ , \qquad 
\left[ H_{S^5} , Q_2^{\rm string}(\lambda') \right]   \neq 0\ .
\ee
There is no charge $Q_2^{\rm string}(\lambda')$ in the 
string theory that satisfies all of the requirements set forth by 
the gauge theory.  This of course indicates the breakdown of integrability 
at $O(1/J^2)$ in the curvature expansion.  

This result, however, can also be seen to 
indicate a larger inconsistency within the string analysis.
At zeroth order in $\lambda'$ the eigenstates 
$a_s^{\dag (a'}a_r^{\dag b'}a_q^{\dag c' )} \Ket{J}$
and $a_s^{\dag (a}a_r^{\dag b}a_q^{\dag c )} \Ket{J}$ can be reinterpreted
as zero-mode string states and the computation above amounts to an 
energy eigenvalue calculation for the superparticle.  From
the analysis put forth in \cite{Callan:2004uv}, the superparticle
spectrum cannot acquire any corrections associated with the curvature
of the target space geometry.  In this light, eqn.~(\ref{AdSMX2})
states that the string theory fails to meet a fairly basic constraint.

\section{Discussion and conclusions}

Previous tests of the AdS/CFT correspondence near the BMN limit
have indicated that, at $O(1/J)$, the string theory begins to disagree
with gauge theory predictions at three loops in the gauge coupling.  
Originally, this disagreement seemed to signal a
failure of string theory integrability at that order, since
the $SU(2)$ SYM dilatation operator is uniquely fixed by assuming 
integrability (and proper BMN scaling) 
\cite{Beisert:2003tq,Beisert:2003jb}.  
Following the study in \cite{3IMP}, however, it
became apparent that the string Hamiltonian at $O(1/J)$ preserves
parity degeneracy among three-impurity string states and commutes
with $Q_2^{\rm string}(\lambda')$.  While it may have been promising 
that this sector of the integrable structure is preserved at $O(1/J)$, 
it now appears that, if integrability is to survive at $O(1/J^2)$, 
some additional ingredient is needed.

It has been suggested that higher-order disagreements with gauge theory may be due 
to an order-of-limits issue \cite{Serban:2004jf}.  Specifically, we assume
that there is some expression for a given charge on either side of
the duality that is exact in $\lambda'$ and $J$ (or $\lambda$ and $R$):
$Q_k^{\rm string}(\lambda',J) = Q_k(\lambda,R)$.  On the string side, 
the charges $Q_k^{\rm string}(\lambda',J)$ are first expanded in powers of $1/J$, 
followed by an expansion in $\lambda'$ for comparison with the gauge theory.
Conversely, the gauge theory (spin chain) charges $Q_k(\lambda,R)$ are derived perturbatively
near $\lambda = 0$, and a subsequent expansion in the 
$R$-charge (or spin chain length $L$) is performed for comparison with the string side. 
In \cite{Serban:2004jf} it was shown that the order in which these limits 
are taken can lead to an erroneous thermodynamic limit ($L\to \infty$) 
in certain spin chain systems (a limit that is naturally well-defined in the string theory).  
While this particular problem is resolved in \cite{Beisert:2004hm}, 
it provides a concrete example of how these issues can lead to superficially
discordant results between both sides of the duality.  A more recent suggestion
involving spin-chain wrapping interactions was given in \cite{Beisert:2004hm}.
Wrapping terms, characterized as having an interaction range greater 
than the length of the spin chain, should naturally affect corrections 
to the BMN limit that specifically incorporate finite-length effects. 
In the gauge theory, the loop expansion followed by the thermodynamic
limit is expected to drop wrapping terms, while the inverse operation 
on the string side is expected to include these effects.
At present these considerations have not been realized in any quantitative fashion.  
To include these effects in the spin chain 
analysis, one would have to sum all perturbative loop corrections prior to 
taking the thermodynamic limit.  This is a daunting proposal but, 
in light of the recent developments in \cite{Beisert:2004hm}, 
such a computation may soon be within reach.  

At this stage we do not have a precise algorithm for rescuing integrability
in the string theory or interpreting the failure thereof in the context of the 
AdS/CFT correspondence.  If the string spectrum is to remain internally consistent
at zeroth-order in $\lambda'$, however, it must also align with integrability 
expectations at that order.  One possibility is that the small-$\lambda'$ expansion
should not be executed before computing corrections associated with second-order
intermediate state sums mediated by $H^{(2)}$.  With our current computing capabilities,
however, this has not yet been possible.
When this problem is solved and higher-order $\lambda'$ 
corrections to the string spectrum are successfully computed at $O(1/J^2)$,  the methods developed 
here will provide a simple and concrete test of integrability and
of any mechanism that hopes to resolve the standing
mismatch between string and gauge theory at three-loop order in $\lambda'$.

\section*{Acknowledgments}

I would like to thank Curtis Callan and John Schwarz for
reading the manuscript.
I am especially indebted to Niklas Beisert and Tristan McLoughlin
for helpful suggestions and interesting discussions. 
I also thank
Gleb Arutjunov,
Charlotte Kristjansen,
Martin Kruczenski,
Didina Serban and
Matthias Staudacher 
for useful comments
and for pointing out references.  
This work was 
supported in part by US Department of Energy grant DE-FG03-92-ER40701. 




\end{document}